# All-dielectric Structure for Trapping Nanoparticles via Light Funneling and Nanofocusing

Amir M. Jazayeri[1,*] and Khashayar Mehrany[1]

[1]Department of Electrical Engineering, Sharif University of Technology, Tehran 145888-9694, Iran
*Corresponding author: jazayeri@ee.sharif.edu

We propose a dielectric structure which focuses the laser light well beyond the diffraction limit and thus considerably enhances the exerted optical trapping force upon dielectric nanoparticles. Although the structure supports a Fabry-Perot resonance, it actually acts as a nanoantenna in that the role of the resonance is to funnel the laser light into the structure. In comparison with the lens illuminating the structure, the proposed structure offers roughly a 10000-fold enhancement in the trapping force – part of this enhancement comes from an 80-fold enhancement in the field intensity, and the remaining comes from a 130-fold enhancement in the normalized gradient of the field intensity (viz. the gradient of the field intensity divided by the field intensity). Also, the proposed structure offers roughly a 100-fold enhancement in the depth of the trapping potential. It is noteworthy that 'self-induced back-action trapping' (SIBA), which has recently been the focus of interest in the context of optical resonators, does not take place in the proposed resonator. In this paper, we also point out some misconceptions about SIBA together with some hitherto unappreciated subtleties of the dipole approximation.

## 1. INTRODUCTION

Dielectric microparticles can be trapped and manipulated noninvasively via light [1,2]. The necessity of having a trapping force stems from the inevitability of the thermal motion of particles. The trapping force is sometimes referred to as the gradient force in that it pulls particles toward a point where the field intensity (viz. the electric field squared) is locally maximum. The idea of trapping particles cannot be easily extended to include dielectric nanoparticles in that the exerted gradient force upon a dielectric particle scales with its volume [3].

For a given laser power and a given particle size, the trapping force can be enhanced by increasing the field intensity or its normalized gradient (viz. the gradient of the field intensity divided by the field intensity). The field intensity and its normalized gradient are both increased by confining light beyond the diffraction limit. Usually plasmonic structures are employed to confine light beyond the diffraction limit and thus to realize three-dimensional trapping of nanoparticles [4-6].

Since plasmonic structures are afflicted by the Ohmic loss and its consequences, the main purpose of this paper is to propose a dielectric structure for three-dimensional trapping of nanoparticles. To this end, one idea is to employ a dielectric resonator, and thus to enhance the field intensity. The dielectric resonators proposed so far are few, and are comprised mainly of photonic crystal cavities [7-9]. In contrast, plasmonic resonators account for the vast majority of the resonators employed so far [10,11] in that they enjoy subwavelength spatial features.

Although the resonator we propose in this paper is dielectric, it supports subwavelength spatial features.

Interestingly, in most of the dielectric and plasmonic resonators proposed so far, neither the field intensity nor its normalized gradient is regarded as the principal factor in trapping particles [8-11]. Rather, 'self-induced trapping' has been touted as the principal trapping mechanism. Self-induced trapping, which was coined in [8] and subsequently called 'self-induced back-action trapping' (SIBA) in [12], refers to the situation where the presence of the particle to be trapped causes a significant shift (viz. a shift comparable to or larger than the linewidth of the resonator) in the resonance frequency. By definition, SIBA requires the linewidth of the resonator to be smaller than a threshold proportional to the particle size. Since the lifetime of photons in the state-of-the-art resonators is hardly longer than a few nanoseconds, SIBA cannot take place when the particle size decreases to a few tens of nanometers. On the other hand, implicit in SIBA is the assumption that the linewidth of the resonator is so large that the resonator responds to the particle's motion instantaneously [13]. Also, the laser frequency must be detuned from the bare resonance frequency (viz. from the resonance frequency in the absence of the particle) so that the detuning *compensates* for the significant resonance frequency shift induced by the presence of the particle.

The term 'SIBA', in our opinion, is a misnomer in that the resonance frequency $\omega_R$ is *always* sensitive to the position of the particle whenever the electromagnetic (EM) mode corresponding to $\omega_R$ contributes to the exerted force upon the particle. Also, as we discuss in Appendix A, we believe that it is a misconception that SIBA increases the trapping force and the magnitude of the trapping potential (viz. the line integral of the trapping force when the particle is moved from its position to infinity). In fact, the trapping force and the magnitude of the trapping potential in the situation where SIBA does not take place (viz. when the particle does not significantly change the resonance frequency, and, the laser is not detuned from the resonance frequency) are larger than (or equal to) the corresponding values in the situation where SIBA takes place. However, SIBA does change the trapping potential profile. For instance, one may increase the width of the trapping potential profile by choosing an appropriate value for the detuning [13]. However, it should be noted that such an increase comes at the expense of a decrease in the trapping force and the depth of the trapping potential.

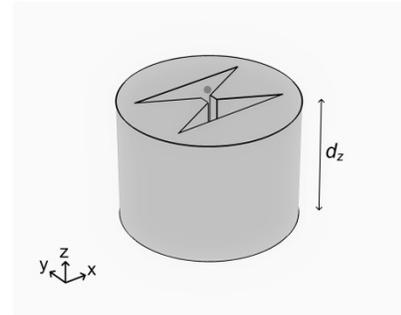

**Fig. 1.** A schematic diagram of the proposed structure, which comprises a bowtie aperture bored in a dielectric of high refractive index, e.g. silicon. The laser light whose electric field is linearly polarized along the *x* axis propagates along the *z* axis, and illuminates the lower end of the structure, while, as is shown in the figure, the nanoparticle is to be trapped around the upper end of the structure. The height of the structure is denoted by $d_z$. Other geometrical parameters of the structure are defined in Fig. 2(a). We assume that the origin of the coordinate system lies at the center of the upper end of the structure.

## 2. PROPOSED STRUCTURE

The dielectric structure we propose in this paper and the plasmonic structure proposed in [14] are two examples of the resonator-based optical tweezers in which SIBA does not take place. Our proposed structure, which is schematically shown in Fig. 1, is a bowtie aperture bored in a cylindrical dielectric of high refractive index (e.g. silicon). We could also choose a non-circular cross-section for the dielectric (e.g. a square cross-section). Once the laser light which is linearly polarized along the *x* axis illuminates one end of the structure, a trapping force appears in the vicinity of the other end. The necessity of boring a bowtie aperture in a dielectric of high refractive index, finite thickness, and finite height can be understood by considering the following facts. First, EM fields can be confined within a slot (viz. a *non-square* rectangular aperture) bored in a dielectric of high refractive index and finite thickness. This one-dimensional field confinement within the region of low refractive index can be attributed to the continuity of the normal component of the electric displacement field [15,16]. Second, the desired two-dimensional field confinement, which is depicted in Fig. 2(a), is realized by reshaping the slot into a bowtie aperture so that the effective refractive index gradually increases as we move along the *y* axis toward the axis of *z* axis. Third, this two-dimensional field confinement leads to the *x* and *y* components of the trapping force while diffraction plays an important role in the *z* component of the trapping force. Fourth, a Fabry-Perot resonance along the *z* axis ensures that the laser light funnels into the aperture. The transmittance spectrum of the structure, which is defined as the ratio of the power exiting the *aperture* to the power incident on the *aperture* versus the light wavelength, is plotted in Fig. 2(b). A transmittance larger than unity indicates light funneling into the *aperture*. Numerical simulations show that the presence of a particle of refractive index 2 and diameter 10 nm has virtually no effect on the transmittance spectrum even if the particle enters the aperture. The same is true for a structure with a smaller value for *w* (e.g. with *w*=15 nm). Therefore, SIBA does not take place in our proposed resonator. The results of the force calculations, which are presented in Section 4, will also confirm that SIBA does not take place.

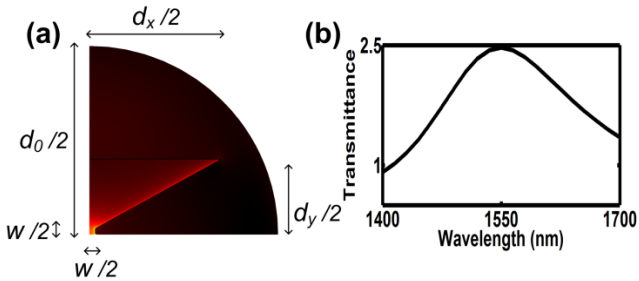

**Fig. 2.** (a) One quarter of a cross-section of the proposed structure schematically shown in Fig. 1. The color represents the electric field profile on upper end of the structure when the laser light which is linearly polarized along the *x* axis illuminates the lower end of the structure. The light wavelength in free space is 1550 nm, and thus the structure, which is made of silicon, is loss-less. The lower end of the cylinder is placed at the focal plane of a conventional lens whose numerical aperture is 0.8. The geometrical parameter values of the structure are $d_0$=500 nm, $d_x$=350 nm, $d_y$=200 nm, $d_z$=360 nm, and *w*=30 nm. (b) The transmittance spectrum of the structure, which is defined as the ratio of the power exiting the *aperture* to the power incident on the *aperture* versus the light wavelength.

A transmittance larger than unity indicates light funneling into the *aperture*. The calculated transmittance would be much larger if transmittance was defined as the ratio of the power exiting the central part of the aperture to the power incident on this part.

## 3. EM FORCE CALCULATION

The Maxwell stress tensor is employed to calculate the time-averaged EM force exerted upon the particle to be trapped:

$$\vec{F} = \frac{1}{2}\text{Re}\left\{\oint_S d\vec{S} \cdot \left[\varepsilon_0 \vec{E}^* \vec{E} + \frac{1}{\mu_0}\vec{B}^* \vec{B} - \frac{1}{2}\vec{I}(\varepsilon_0 \vec{E}^* \cdot \vec{E} + \frac{1}{\mu_0}\vec{B}^* \cdot \vec{B})\right]\right\}, \quad (1)$$

where $\vec{E}$ is the electric field phasor, $\vec{B}$ is the magnetic field phasor, $\vec{I}$ is the identity tensor, $\varepsilon_0$ and $\mu_0$ are the permittivity and permeability of free space, respectively, and $S$ is any closed surface which does not enclose any material body other than the particle [17].

The form of the Maxwell stress tensor used in Eq. (1) comes from the Ampere-Lorentz formulation of the EM force density [17]. The Minkowski and Einstein-Laub formulations are among other formulations of the EM force density. Each formulation leads to a distinct form of the Maxwell stress tensor, and also to a distinct expression for the momentum density of light inside matter. The Minkowski formulation leads to the Minkowski momentum density, whereas the Einstein-Laub formulation leads to the Abraham momentum density. The Ampere-Lorentz formulation leads to the Livens momentum density, which equals the Abraham momentum density in non-magnetic media. It has been shown that the Einstein-Laub formulation is incompatible with special relativity [17]. Some have argued that Minkowski's is the correct formulation [18], while some others have argued that it is not [19]. Nonetheless, all formulations of the EM force density lead to the same result for the EM force (or time-averaged EM force) exerted upon an object *if* EM fields are static (or time-harmonic) *and* the object is immediately surrounded by free space. Since both of these conditions are met in the numerical examples presented in Section 4, any other form of the Maxwell stress tensor could be used in Eq. (1). Since EM fields are time-harmonic, 'time-averaged EM force' will hereafter be shortened to 'EM force' in the paper.

Since the size of the particle is much smaller than the light wavelength, the dipole approximation *may* be applicable. Under the dipole approximation, the particle is replaced with a linearly polarizable point-like dipole [20]. By applying the dipole approximation to Eq. (1), the z component of the EM force is simplified to [21]:

$$F_{\text{dip}_z} = \frac{\alpha_R}{4}\partial(\vec{E}_i^* \cdot \vec{E}_i)/\partial z + \frac{\alpha_I}{2}\text{Im}[\vec{E}_i^* \cdot (\partial \vec{E}_i / \partial z)], \quad (2)$$

where $\vec{E}_i$ is the *incident* electric field phasor (viz. the electric field phasor in the *absence* of the particle), $\alpha_R$ and $\alpha_I$ are the real and imaginary parts of the polarizability of the particle, respectively, and the derivatives are calculated at the center of the particle. Other terms would be added to Eq. (2) if the particle had a magnetic response besides its electric response [22]. The EM force along other coordinate axes can be written in a similar fashion. The first term on the right-hand side of Eq. (2) is the z component of the gradient force while the second term is the z component of radiation pressure. Interestingly, as we show in Appendix B, the gradient force in Eq. (2) can also be derived by applying the dipole approximation to the method of virtual work instead of the Maxwell stress tensor.

Assuming that the particle is spherical and *low-loss*, its polarizability reads [23]:

$$\alpha = 4\pi\varepsilon_0 R^3 \alpha_0 / [1 + j2(k_0 R)^3 \alpha_0 / 3], \quad (3)$$

where $k_0$ is the wavenumber in free space, $R$ is the radius of the particle, and $\alpha_0$ denotes $(\varepsilon - 1)/(\varepsilon + 2)$ in terms of its relative permittivity. Although $\alpha_I$ never vanishes, it becomes negligible when the particle is loss-less [viz. when $\text{Im}(\varepsilon)$ is zero]. Since the particle we consider in the numerical examples is loss-less, we ignore radiation pressure whenever the results of Eq. (2) are presented. Under the dipole approximation, the trapping potential is readily found to be $-0.25\alpha_R \vec{E}_i^* \cdot \vec{E}_i$.

The EM modes excited by the laser all contribute to the incident electric field in Eq. (2). If $\vec{E}_{\text{res,i}}$ denotes the electric field phasor of one the EM modes in the *absence* of the particle, Eq. (2) states that the z component of the gradient force coming solely from the EM mode reads $F_{\text{gr}_z} = 0.25\alpha_R \partial(\vec{E}_{\text{res,i}}^* \cdot \vec{E}_{\text{res,i}})/\partial z$. Assuming that SIBA happens to this EM mode, the presence of the particle causes a significant change in the *amplitude* of the EM mode (or, equivalently, the photon occupancy in the EM mode), and therefore, $F_{\text{gr}_z}$ is no longer equal to the z component of the actual EM force the EM mode exerts upon the particle. Therefore, if the EM mode is intended to make a major contribution to the total EM force exerted upon the particle, a significant discrepancy between Eq. (2) and Eq. (1) appears when SIBA happens to the EM mode.

It should be noted that SIBA is only one of the situations in which the dipole approximation fails. SIBA merely refers to the situation where the presence of the particle significantly changes the *amplitude* of one of the EM modes of the system but does not significantly change its electric field *profile* [13]. In fact, whenever the presence of the particle significantly changes the amplitude(s) or field profile(s) of some of the EM modes of the system, or introduces new EM modes into the system, the dipole approximation fails. In general, the failure of the dipole approximation requires at least one of the following two conditions to be met: (i) the particle is plasmonic [24] or of high refractive index [25], and (ii) the incident electric field has strong spatial variations over length scales comparable to the particle size (although the wavelength of the laser may be much larger than the particle size). It should be noted that when the dipole approximation fails, the gradient force under the dipole approximation may be smaller or larger than the actual EM force. As we see in Section 4, our proposed structure exemplifies this point.

Also, it should be noted that the trapping potential is not well-defined when the gradient force under the dipole approximation and the actual EM force disagree, in that the EM force is no longer conservative. In such a case, the path over which the trapping force is integrated must be specified.

## 4. RESULTS AND DISCUSSION

We use the finite element method (COMSOL Multiphysics) to numerically calculate $\vec{E}$ and $\vec{B}$ in Eq. (1) as well as $\vec{E}_i$ in Eq. (2). We make the following assumptions besides the assumptions made in Fig. 1 and Fig. 2(a): (i) the particle to be trapped is a loss-less sphere of refractive index 2 and diameter 10 nm, (ii) the position of the particle is defined as the position of its center with respect to the center of the upper end of the structure, and (iii) the total laser power is 1 mW, and thus, the calculated forces are normalized to 1 mW of the total laser power. It is worth noting that the power funneling into the aperture is maximized for $d_z$=356 nm, whereas the trapping force is maximized for $d_z$=364 nm in that the trapping force depends not only on the power funneling into the aperture but also on the gradient of the field intensity. However, $d_z$ can be changed in the range from 345 nm to 382 nm and the trapping force drops only by 10% from its optimum value.

The $x$, $y$, and $z$ components of the trapping force versus $x$ on ($y$=0, $z$=10nm), versus $y$ on ($x$=0, $z$=10nm), and versus $z$ on the $z$ axis are plotted in Figs. 3(a)-(c), respectively. In these figures, the solid lines are the exact numerical results [viz. Eq. (1)] while the dashed lines are the numerical results under the dipole approximation [viz. the gradient force in Eq. (2) and its counterparts along other coordinate axes].

The $y$ component of the trapping force vanishes on ($y$=0, $z$=10nm), but the $z$ component does not. As we depart from $x$=0 on ($y$=0, $z$=10nm), the $x$ component grows from zero to its maximum value, and becomes of the same order of magnitude as the $z$ component. Ultimately, the components both disappear at $x$=∞. The reason for choosing the $x$ component to be plotted on ($y$=0, $z$=10nm) is that the $z$ component does not contribute to the trapping potential if the trapping force is integrated over this line. A similar discussion applies to the trapping force on ($x$=0, $z$=10nm). Also, the $x$ and $y$ components both vanish on the $z$ axis.

According to Fig. 3(c), the dipole approximation is valid as the particle is moved on the $z$ axis even if the particle enters the aperture. It confirms that SIBA does not take place. However, according to Fig. 3(a), the dipole approximation occasionally fails as the particle is moved on ($y$=0, $z$=10nm). The occasional failure of the dipole approximation is due to strong spatial variations in the incident electric field. Interestingly, the maximum value of the $x$ component of the gradient force under the dipole approximation is 32% *larger* than the actual maximum value in Fig. 3(a), whereas it would be 10% *smaller* than the actual maximum value if the particle size was tripled, and the particle was moved on ($y$=0, $z$=20nm). It demonstrates that when the dipole approximation fails, the actual EM force may be smaller or larger than the gradient force under the dipole approximation.

Since the dipole approximation occasionally fails in our proposed structure, the path over which we integrate the trapping force to calculate the trapping potential must be specified. The points at which the trapping potential will be reported all lie on the $z$ axis. One of these points is the trapping point, where the three components of the trapping force all vanish. The path over which the trapping potential is integrated is the $z$ axis.

According to the exact numerical results (viz. the solid lines) in Fig. 3, the maximum values of the $x$ and $y$ components of the trapping force on ($y$=0, $z$=10nm) and ($x$=0, $z$=10nm) are 1.0 fN and 0.7 fN, respectively. The $z$ component of the trapping force at ($x$=0, $y$=0, $z$=10nm) is 2.1 fN. The maximum value of the $z$ component of the trapping force on the $z$ axis is 3.0 fN. The trapping point is ($x$=0, $y$=0, $z$=−21 nm). The trapping potential at the trapping point and at ($x$=0, $y$=0, $z$=10nm) is -0.021$k_B T$ and -0.008$k_B T$, respectively, where $k_B$ is the Boltzmann constant, and $T$ is 293 K.

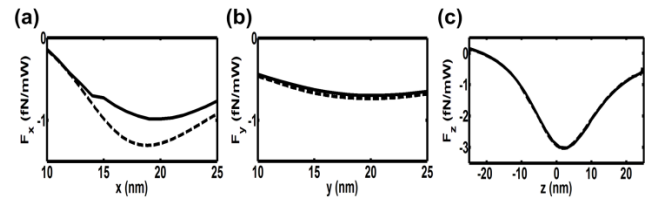

**Fig. 3.** The trapping force (in fN) normalized to 1 mW of the total laser power versus the center-of-mass position (in nm) of a loss-less nanoparticle of refractive index 2 and diameter 10 nm. The nanoparticle is to be trapped around the upper end of the structure schematically shown in Fig. 1. The position of the particle is defined as the position of its center with respect to the center of the upper end of the structure. All other assumptions are identical to those made in Fig. 2(a). The solid lines are the exact numerical results while the dashed lines are the numerical results under the dipole approximation. (a) The $x$ component versus $x$ on ($y$=0, $z$=10nm), where the $y$ component vanishes. (b) The $y$ component versus $y$ on ($x$=0, $z$=10nm), where the $x$ component vanishes. (c) The $z$ component versus $z$ on the $z$ axis, where the $x$ and $y$ components both vanish.

The trapping force and the magnitude of the trapping potential can be increased by decreasing *w*. In the second set of numerical examples, *w* is halved. The *x*, *y*, and *z* components of the trapping force versus *x* on (*y*=0, *z*=10nm), versus *y* on (*x*=0, *z*=10nm), and versus *z* on the *z* axis are plotted in Figs. 4(a)-(c). The maximum values of the *x* and *y* components of the trapping force on (*y*=0, *z*=10nm) and (*x*=0, *z*=10nm) are now both 1.2 fN. The *z* component of the trapping force at (*x*=0, *y*=0, *z*=10nm) is now 4.4 fN. More importantly, the maximum value of the *z* component of the trapping force on the *z* axis is now 15.7 fN (viz. more than 5 times larger than the corresponding value when *w* is 30 nm). The trapping point is again (*x*=0, *y*=0, *z*= −21 nm). The trapping potential at the trapping point and at (*x*=0, *y*=0, *z*=10nm) is now -0.054$k_BT$ and -0.01$k_BT$, respectively.

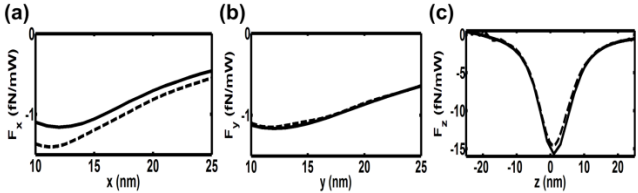

**Fig. 4.** The trapping force (in fN) normalized to 1 mW of the total laser power versus the center-of-mass position (in nm) of a loss-less nanoparticle of refractive index 2 and diameter 10 nm. All assumptions are identical to those made in Fig. 3 except that *w* is now 15 nm.

For a given value of *w*, one could further increase the trapping force and the magnitude of the trapping potential by optimizing the values of $d_0$, $d_x$, $d_y$, and $d_z$. However, the parameter values used in the numerical examples presented above are not optimized.

It is emphasized that the light wavelength in free space is 1550 nm, whereas the optical tweezers proposed in the recent literature usually enjoy smaller light wavelengths. Also, the calculated forces presented above are normalized to 1 mW of the total laser power, whereas the forces reported in the literature are not necessarily normalized to the total laser power. For instance, in [14], the light wavelength in free space is 692 nm, and the calculated forces are normalized to the *transmitted* optical power, which can be much smaller than the required total laser power.

To investigate how much enhancement in the trapping force is achieved with our proposed structure, we compare the trapping force in the presence of the structure against the gradient force the lens (viz. the same lens illuminating the structure) would exert on the particle in the absence of the structure. Since the electric field profile of the laser light exiting the lens is Gaussian [26], and the size of the particle is much smaller than the light wavelength, the gradient force in the absence of the structure can be calculated analytically by using Eq. (2). Such a calculation shows that the maximum value of the *z* component of the gradient force normalized to the total laser power is equal to $0.3NA^4(k_0R)^3\alpha_0/c$ in the absence of the structure, where *c* is the speed of light in free space, *NA* is the numerical aperture of the lens, and $k_0$, *R*, and $\alpha_0$ retain their previous definitions in Eq. (3). Therefore, according to Fig. 3(c), there is roughly a 2000-fold enhancement in the *z* component of the trapping force when the structure with *w*=30 nm is employed – part of this 2000-fold enhancement comes from a 30-fold enhancement in the field intensity, while the remaining comes from a 70-fold enhancement in the normalized gradient of the field intensity (viz. in the gradient of the field intensity divided by the field intensity). Also, according to Fig. 4(c), there is roughly a 10000-fold enhancement in the *z* component of the trapping force when the structure with *w*=15 nm is employed – part of this 10000-fold enhancement comes from an 80-fold enhancement in the field intensity, while the remaining comes from a 130-fold enhancement in the normalized gradient of the field intensity. One might also like to investigate the enhancements in the *x* and *y* components of the trapping force, whose maximum values normalized to the total laser power are both equal to $0.6NA^3(k_0R)^3\alpha_0/c$ in the absence of the structure. In such a case, it should be noted that the maximum values of the *x* and *y* components of the trapping force in the presence of the structure are larger than the maximum values deduced from Figs. 3(a),(b) and Figs. 4(a),(b) in that these figures are plotted at a specific *z* (viz. at *z*=10 nm).

To investigate how much enhancement in the depth of the trapping potential (viz. the magnitude of the trapping potential at the trapping point) is achieved with our proposed structure, it suffices to compare the field intensity at the trapping point in the presence of the structure against the field intensity at the focal point of the lens in the absence of the structure. There is roughly a 40-fold enhancement in the depth of the trapping potential when the structure with $w$=30 nm is employed, and a 100-fold enhancement when the structure with $w$=15 nm is employed.

A trapping potential of depth $10k_BT$ ($k_B$ is the Boltzmann constant, and $T$ is 293 K) for a particle of refractive index 2 and diameter 10 nm illuminated at the free space wavelength 1550 nm requires a total laser power of 185 mW, and 476 mW when the structure width is $w$=15 nm, and $w$=30 nm, respectively. The same trapping potential for the same particle requires a total laser power of 18 W when only a lens of numerical aperture 0.8 (viz. the same lens illuminating the structure) is employed instead of the proposed structure. Moreover, the trapping potential of depth $10k_BT$ is accompanied by a trapping force whose maximum value on the $z$ axis is 2.9 pN, and 1.4 pN when the structure width is $w$=15 nm, and $w$=30 nm, respectively. The trapping force accompanying the same trapping potential for the same particle reaches a maximum value of 30 fN on the $z$ axis when the structure is substituted with a lens of numerical aperture 0.8. Finally, it should be noted that if we attribute a dielectric loss to the particle, the particle reaches the same temperature at the trapping point (where the magnitude of the trapping potential is $10k_BT$) whether the structure width is $w$=15 nm or $w$=30 nm, or the structure is substituted by a lens.

## APPENDIX A

Here, we argue that SIBA does not increase the trapping force or the magnitude of the trapping potential. By looking into the Hamiltonian describing the interaction between a particle and an EM mode of a resonator, and ignoring quantum fluctuations, the $z$ component of the trapping force coming (solely) from the EM mode can be written as $F_z(\vec{r}) = -\hbar N(\vec{r}) \partial \omega_R(\vec{r}) / \partial z$, where $\hbar$ is the reduced Planck constant, and the mean intracavity photon number ($N$) and the resonance frequency ($\omega_R$) of the EM mode are both functions of the position of the particle ($\vec{r}$). On the other hand, implicit in SIBA is the assumption that the presence of the particle does not significantly change the electric field *profile* of the EM mode [12], although it changes the *amplitude* of the EM mode (or, equivalently, the photon occupancy in the EM mode). This assumption allows one to calculate $\omega_R(\vec{r})$ by applying a perturbation method to the EM wave equation. Interestingly, for instance, for a Fabry-Perot resonator, it can be seen that the calculated $F_z(\vec{r})/N(\vec{r})$ is equal to $F_{g_z}(\vec{r})/N_g$, where $N_g$ denotes the mean intracavity photon number in the *absence* of the particle, and $F_{g_z}(\vec{r})$ is the $z$ component of the gradient force derived by applying the dipole approximation (viz. the approximation that the presence of the particle has the least possible effect on EM fields) to the Maxwell stress tensor [see Eq. (2)]. Therefore, $F_z(\vec{r})$ can be written as $[N(\vec{r})/N_g]F_{g_z}(\vec{r})$. It should be noted that, when SIBA takes place, the laser frequency must be detuned from the bare resonance frequency (viz. from the resonance frequency in the absence of the particle). Evidently, $F_z(\vec{r})$, $N(\vec{r})$, $F_{g_z}(\vec{r})$, and $N_g$ are all defined in the presence of such a detuning.

Similarly, if SIBA did not take place (viz. if the presence of the particle did not significantly change the resonance frequency, and, the laser was not detuned from the resonance frequency), the trapping force would equal $[N_0/N_g]F_{g_z}(\vec{r})$, where $N_0$ is the mean intracavity photon number in the *absence* of the particle and detuning *both*. Since $N(\vec{r})$ never exceeds $N_0$, the trapping force in the situation where SIBA takes place never exceeds the trapping force in the situation where SIBA does not take place. In other words, SIBA does not increase the trapping force at *any* point. As a result, SIBA does not increase the magnitude of the trapping potential at any point either. It should be noted that, as we discuss in Section 3, the trapping potential is not well-defined when the dipole approximation fails. Therefore, when SIBA takes place, the path over which the trapping force is integrated must be specified.

## APPENDIX B

Here, we show that the gradient force in Eq. (2) can also be derived by applying the dipole approximation to the method of virtual work instead of the Maxwell stress tensor. Under the dipole approximation, the presence of a small non-magnetic particle does not significantly change the magnetic field. Therefore, the magnetic energy and the curl of the electric field are both almost insensitive to the position of the particle. The insensitivity of the magnetic energy to the position of the particle allows us to exclude it from the calculation of virtual work. The insensitivity of the curl of the electric field to the position of the particle allows us to make use of the electrostatic result derived in [27]. The electrostatic result states that the $z$ component of the exerted electric force upon a linearly polarizable dipole $\vec{\mathscr{P}}$ placed in an *initial* electrostatic field $\vec{\mathscr{E}}_i$ whose sources are *fixed* is given by $0.5\partial(\vec{\mathscr{P}} \cdot \vec{\mathscr{E}}_i)/\partial z$. Therefore, since our electric field is actually time-harmonic, the $z$ component of the time-averaged force becomes $0.25\,\mathrm{Re}[\partial(\vec{P}^* \cdot \vec{E}_i)/\partial z]$, which equals $0.25\alpha_R \partial(\vec{E}_i^* \cdot \vec{E}_i)/\partial z$, and corroborates the gradient force in Eq. (2).

**Acknowledgement.** Amir M. Jazayeri is deeply indebted to Dr Farhad A. Namin for providing the computer equipment used to carry out the numerical simulations.